# The entrepreneurial role of the University: a link analysis of York Science Park

David Minguillo and Mike Thelwall

d.minguillobrehaut@wlv.ac.uk
Statistical Cybermetrics Research Group, School of Computer Science and Information Technology,
University of Wolverhampton, Wolverhampton (UK)

**Abstract**
This study introduces a structured analysis of science parks as arenas designed to stimulate institutional collaboration and the commercialization of academic knowledge and technology, and the promotion of social welfare. A framework for key actors and their potential behaviour in this context is introduced based on the Triple Helix (TH) model and related literature. A link analysis was conducted to build an inter-linking network that may map the infrastructure support network through the online interactions of the organisations involved in York Science Park. A comparison between the framework and the diagram shows that the framework can be used to identify most of the actors and assess their interconnections. The web patterns found correspond to previous evaluations based on traditional indicators and suggest that the network, which is developed to foster and support innovation, arises from the functional cooperation between the University of York and regional authorities, which both serve as the major driving forces in the trilateral linkages and the development of an innovation infrastructure.

**Introduction**

The necessity to translate academic knowledge into technological applications as a source of social and economic development has transformed the traditional teaching and research role of the university. This new social mission is turning academia into the central actor of a complex network in which interactions with industrial and governmental spheres are fundamental to commercialize research and create an organizational system of innovation (Etzkowitz, 2006; Godin & Gyngras, 2000). In this context, government support can be seen in the adoption of programs and policies which promote and strengthen mutual dependence between research institutions and enterprises, leading to the decline of institutional boundaries and the emergence of a triadic sub-network of actors which work together for a common benefit (Etzkowitz, 2008). The growth of cross-sectoral interactions is promoted by science, technology and innovation policies which create organizational mechanisms for innovation. Science parks (SPs) are an example of this and these hybrid organizations adopt various roles to bring academia closer to industry by promoting the commercialization of academic research. SPs are expected to exploit the research strength of a university in particular areas to foster the creation of university spin-offs and strategic alliances. Thus, SPs are also designed to attract new technology-based firms and the R&D units of existing companies that desire collaboration with academic researchers (Etzkowitz, 2008; Vedovello, 1997).

However, despite significant academic and public expenditure in SPs their economic value is still unproven (Löfsten & Lindelöf, 2002; Quintas, Wield, & Massey, 1992), and there is a clear need to investigate their intermediary role in supporting the commercialization of academic knowledge and technology as well as the promotion of social welfare (Suvinen, Konttinen, & Nieminen, 2010). There is also a need to broaden the indicators used to assess the entrepreneurial and social function of universities (Larsen, 2010) and hence, fill the gap in the traditional indicators to evaluate the inherent multidimensional and synergetic nature of TH linkages (Etzkowitz, 2008). A previous attempt to investigate the infrastructure of SPs using link analysis (Minguillo & Thelwall, submitted) shows strong TH cooperation and identifies universities as central in the networks, but in order to gain deeper insights into the configuration of SPs it is necessary to carry out a systematic and structured analysis. Therefore, the purpose of this study is twofold: (1) to design a new framework, the *SP actor framework*, based on the TH model that lists the key actors that should be involved in the SPs and identifies their missions, functions, and potential interactions, and (2) to compare this framework with the network created by the hyperlinks among the actors within the SP to determine if the links reveal potential offline behaviours and whether the identified patterns are in line with the results obtained with traditional indicators. The latter makes sense because hyperlinks can reflect social interactions and represent an underlying social structure (Reid, 2003). These two objectives are addressed by the following research questions:

1. Can the organisations in the *SP actor framework* be identified in a hyperlink network generated by the web sites of organisations associated with a SP?

2. Do the links between organisations in hyperlink network reflect the potential behaviour of the different types of actors described in the *SP actor framework*?

**The SP Actor Framework**

A number of different types of actors, listed below, are often found in organizational innovative environments (Etzkowitz, 2008) although this list does not include the full range of actors that could be present (Howells, 2006). The presence or absence of particular actors in a SP depends on its economic development strategy, which is determined by the local and regional conditions and objectives, as well as the degree of cooperation between the institutional spheres (University-Government-Industry). Thus, the effectiveness of a SP with regards to the commercialization of research could not only be related to the establishment of certain intermediary organizations but also to the functionality and productivity of the interconnections in the network, as highlighted by Suvinen, Konttinen and Nieminen (2010). The collaboration within the network may facilitate the hybridisation needed to overcome the absence of certain types of actors, so the remaining actors, especially intermediaries, would take on the missing roles in order to fill the gaps in the innovation infrastructure (Etzkowitz & Leydesdorff, 2000). The intention with this framework is therefore not only to identify the key social actors and their behaviours but also to learn which functions and roles are important to create an innovative environment and examine which actors could fill these roles.

*University:* As an institution, the university interacts with industry and government to embrace a new social mission that expands its academic and scientific missions. It attempts to commercialize its research by connecting to industry, as a potential knowledge consumer, and thereby attracts external funding, promotes employment and fosters regional strategic development. This entrepreneurial function requires the development of technology transfer capabilities in the form of consulting, patenting and licensing, and firm-formation activities that can lead to independent entities emerging from the university or at least having strong academic links. These firms are often located in dynamic and quasi-academic spaces like SPs. *Role in network:* The university is the driving force in a SP and is expected to occupy a central position in an innovative support structure. The entrepreneurial university keeps strong ties with hybrid actors rooted in academia that produce, capitalize, and disseminate knowledge, such as technology transfer offices, research centres, consulting organizations, incubators. It is also expected to have direct connections with government and industry actors, such as regional development agencies (RDA), R&D units of large firms, start-ups, spin-offs, and other actors that support the development of the network and need research and advanced technology (Etzkowitz, 2008; Etzkowitz, et al., 2000).

***Research centres:*** These partly fulfil the research missions of universities and typically involve strategic alliances to achieve long-term goals. They may bring various intellectual, physical and organizational resources together within a single university or span several universities and non-academic institutions such as government research institutions and firm laboratories to engage in a more intense and interdisciplinary collaboration. A centre may host several research groups around a theme and for several purposes: (1) to attract more funding, (2) to access to better facilities and instruments, and (3) to undertake large-scale projects.
*Role in network*: Research centres liaise with academia, industry, government, and the wider public to provide a space for collaboration between actors with diverse perspectives who are interested in pursuing applied knowledge. This concentration of resources and interests makes research centres important social actors that may serve a heterogeneous network of organisations that need highly specialized scientific knowledge and technology to resolve particular problems and increase their competitive advantage. Therefore, research centres are expected to occupy a central position in the network and to be linked with knowledge producers and consumers (Etzkowitz, 2008).

***Consulting organizations:*** These are liaison offices that convert informal individual consulting into a professional and organized group activity. They identify technological needs in industry or government and put together the necessary resources to provide new customized solutions. Their strong university links suggest that this function could be filled by the university or a quasi-academic actor that connects the university with a group of consumers related to the private and public sector. Such consulting organizations can be either an economic arm of the university that brings external funding to the academic world, i.e. an internalised consulting model, or an independent business, hiring individual scholars for specific projects, i.e. an externalised consulting model.
*Role in network:* These actors organize and strengthen the interactions between researchers or research units, regional organizations and companies through consultation and research contracts. They represent the first steps to capitalize knowledge, and would therefore be expected to have a close interaction with knowledge consumers, while their connections to the university would depend on the degree of independency of the consultant. (Etzkowitz, 2008:95).

***Technology transfer offices:*** These have a dual search mechanism; an internal mechanism to identify and commercialize relevant research and technology produced by the university, and an external mechanism to identify potential markets for it and bring potential customers to the university. These offices are responsible for identifying, patenting, marketing, and licensing intellectual property in order to attract additional research funding via royalties, licensing fees, and research contracts.
*Role in network:* Technology transfer offices are important brokers of the innovation system which value, protect and sell university inventions. This commercial function could be realized by the university itself or an independent actor which may have links with intellectual property attorneys. These actors would usually have strong academic connections and are one of the main intermediaries between the university and a group of technology-based firms (Etzkowitz, 2008:37, 89; Siegel, Waldman, & Link, 2003)

***Incubators***: University-related incubators are sponsored by the university usually in partnership with other interested players. They are independent units intended to commercialize research, technology and intellectual property produced by the university in the form of new firms. This organizational support structure provides the space and added value to assist spin-offs and start-ups with consultation, business services, straightforward funding, subsidized space, access to university facilities, expertise and assistance of researchers and students, networking opportunities, and other services to encourage entrepreneurship and promote academic-industry collaboration. This improved environment should increase the possibilities for survival and growth of the on-site firms.

*Role in network:* The assisting function of incubators in the growth of university spin-offs makes it possible to identify an entrepreneurial sub-network intended to promote new ventures. The role of this actor primarily consists of establishing relationships between the university and entrepreneurs and to create ties to investors as well as public and private hybrid actors in order to support early-stage firms (Barrow, 2001; Etzkowitz, 2002).

***Investors***: Venture capital (VC) is at the heart of the firm-formation process, and SPs with an incubation facility should create an ideal organized environment for investors who can take advantage of innovative ideas and high growth potential ventures in attractive sectors to invest with reduced risks. The injection of funds and resources by investors plays an important role at all stages of the firm-formation process. Early-stage investment provided by business angels and seed capital investors (see below) are the most common form of investment while their involvement in the network is basically related to the particular number of firms in their portfolios.

*Role in network:* Early stage investments are typically provided by public and private actors that are connected to new and high growth potential companies. Despite the importance of these actors in the network, their quantity of links may be low due to the fact that the interactions tend to be intense but informal between investors and companies in the SPs. The intensity of the hyperlinks may exhibit two patterns according to the degree of collaboration between both parties in relation to the stage of development of the firms; (1) strong and informal ties in the case of business angels and early stage ventures which are necessary to get the business off the ground, and (2) weak and formal ties between VCs and growing firms which are necessary to accelerate the growth of businesses (Barrow, 2001:110).

A *business angel* is defined by the European Business Angels Network (2010) as "*a private individual who invests part of his/her personal assets in a start-up and also shares his/her personal business management experience with the entrepreneur*". These are informal suppliers of high risk capital which adopt functions similar to the business incubators, spending time in mentoring early-stage and growing business. The rise of business angel networks (BANs), in which groups of angles cluster together to pool investment and expertise, increases their investment capacity and confidence through a larger collaboration network (Barrow, 2001). On the other hand, *seed capital* is usually offered by the government, universities, and corporations in which the technology of the new firm has been produced. Government and university VC focuses on a social perspective and supports both long-term and research-based projects, or less-favoured fields and less venture capital-intensive regions through funds, research grants, subsidies or indirect loans to promote economic growth. Early-stage funding is intended to fill the gap of private investments which are oriented to later stages, reduced risk and short-term financial returns. A combination of private-public sponsorship may also operate at this stage (Etzkowitz, 2006; 2008:122-36).

***Government agencies:*** These can directly or indirectly encourage an organized TH collaboration through government and quasi-government agencies. They provide a regulatory environment and also act as a public venture capitalist to increase innovation and regional competitiveness (Etzkowitz, 2008). An innovation policy should have the capacity and autonomy to be driven by institutions at local, regional, central, supranational level, such as the *European Regional Development Fund* (ERDF), the *Regional Development Agencies* (RDA), and the *Business Link*, to design adequate infrastructures.

*Role in network:* At a practical level the region is the key space for economic governance and the RDA are the major mechanism to facilitate central policies. They provide financial support to the innovation infrastructures through the establishment of intermediaries and partnerships with the university and key actors to exploit the accumulated research and to support the growth of knowledge-based businesses as significant sources of employment growth and sustainable development (Webb & Collis, 2000).

***Knowledge-based firms:*** These are firms, such as university spin-offs, spin-ins, start-ups, and R&D units, which emerge from or at least are closely associated with a university or another knowledge-producing institution. The formation process of high-tech and knowledge-based firms requires multiple resources and support, and the cross-fertilization between technical and business skills, which is embedded in collaborative relationships that may include other firms in strategic alliances and actors from the university and government. The open innovation process requires an intensive cooperation that is supported and mediated by the innovation network infrastructure.

*Role in network:* The commercialization of research embodied in firms is the engine of innovation strategies. This cluster of firms is expected to be the central economic actors in the interactions occurring through networks across institutional spheres. They may be linked with the university, governmental and private investors, research groups, incubators, and other universities' economic arms (Etzkowitz, 2006; 2008).

***Service-based firms:*** These are firms which tend to operate with an incremental perspective toward product development, utilizing new combinations of existing technologies to solve a problem or provide a service (Etzkowitz, 2008:54). Start-ups and small- and medium-sized firms (SMEs) can be attracted to SPs for the competitive advantage of a prestigious location and a network of potential customers to market their services to. In this group there are also actors such as recruiting, accounting and marketing firms, lawyers, and web-designers.

*Role in network*: Being market-oriented, these firms are likely to establish competitive relationships and neither engage in advanced research nor develop collaborative relationships with the members of the network. This lower degree of reliance on the innovation infrastructure to perform their activities suggests that they could be more isolated than knowledge-based firms.

**Data and Method**

The research questions were addressed with a case study of York Science Park. This SP was chosen due to its location in a region with significant R&D investments undertaken by the higher education sector, especially by the universities of Leeds, Sheffield, and York as well as RDA Yorkshire Forward and the ERDF. These efforts are oriented towards the development of a regional innovation system to support the firm formation and university-industry links as a means to restructure the subsequent decline of traditional industries and bring about economic dynamism in the region (Dabinett & Gore, 2001; Huggins & Johnston, 2009:234-6). Furthermore, York SP provides the largest network among the SPs in Yorkshire and the Humber region (Minguillo & Thelwall, submitted), which makes it relevant to carry out a deeper analysis of its structure.

In order to generate the hyperlink network for York Science Park, this study draws on the same data which was collected in May 2010 in relation to a previous study (Minguillo & Thelwall, submitted) and the following process was used: The external outlinks (i.e., hyperlinks pointing to other web sites) of York Science Park's website were collected with the crawler SocSciBot. The crawler collected 123 unique site outlinks which were then manually examined to eliminate links to generic websites such as search engines, portals, tourist information, social network services and public transport, and to classify them according to a respective sector (Industry, Academia, Government) and category (see Figure 1). The number of relevant websites identified was 107, but after reducing the URLs to their respective domains (e.g., wlv.ac.uk) or sub-domains (e.g., cybermetrics.wlv.ac.uk) to assign each (sub-)domain to an actor and minimise the impact of multiple levels of websites, the number of actors was reduced to 103. Their inlinks were then extracted with Yahoo!, and outlinks with Bing and the web crawler SocSciBot. These three different sources were used to tackle the low overlap in coverage by search engines and minimise the bias and limitations of each tool. The data from both search engines were collected with the help of the free LexiURL Searcher software that allows you to obtain more than 1,000 results for each query (Thelwall, 2008a). 127,397 inlinks were collected through Yahoo! and 50,816 outlinks through Bing (49,592) and SocSciBot (1,224). After the links were reduced to (sub-)domains, the duplicates of each actor were eliminated and the outlinks were combined, we obtained 62,018 inlinks and 36,544 outlinks. The number of outlinks is constituted by 35,331 (96.7%) from Bing, 428 (1.2%) from SocSciBot, and 785 (2.1%) from both tools. The selection of a highly institutionalised and specialised context as a SP, and of three different sources to gather the data are important to minimise the complexity and the poor reliability of web data (Thelwall, 2008b), as well as to increase the accuracy of the structures. Having all the in- and outlinks of the 103 actors makes it possible to construct bidirectional interlinking networks based on the two different (in- and outlinks) data sets as described in Minguillo and Thelwall (submitted). In this study, the two bidirectional interlinking networks were combined to obtain the most robust and complete network, the intensity of the relationships between actors were dichotomised to measure the number of actors that were interconnected rather than the intensity of the connections, and the self-links were eliminated. This resulted in a bidirectional network of 104 actors (including the York SP) that were connected by 378 links, but to facilitate the representation and analysis of this comprehensive network the implicit external links of the SP to the 103 actors were eliminated and only the incoming links to the SP were considered. This eliminated the actors that were only linked by the SP, reducing the number of actors to 75 (72%) and the number of links to 275 (73%). The number of actors from each category and the interconnections between the categories can be observed in Table 1.

**Table 1. Links between the categories found in York Science Park.**

| Actors | Categories | Service-based firm | Know.-based firm | Consultants/IP-TTOs | Business Dev./Invest. | Academia | Support Struc. Org. | Public & Non-Gov. Org. | Government | Science Park | Total - outlinks | Mean |
|---|---|---|---|---|---|---|---|---|---|---|---|---|
| 24 | Service-based firm | 3 | 1 |  | 1 |  | 7 | 2 |  | 3 | 17 | 0.7 |
| 30 | Knowledge-based firm | 1 | 1 |  | 3 | 3 | 3 | 2 | 2 | 2 | 17 | 0.6 |
| 9 | Consultants/IP-TTOs |  |  |  |  | 3 |  |  |  | 1 | 4 | 0.4 |
| 3 | Business Developers/Investors | 7 | 10 |  | 3 | 6 | 9 | 1 | 1 | 2 | 39 | 13.0 |
| 5 | Academia | 4 | 8 | 2 | 5 | 8 | 11 | 11 | 2 | 2 | 53 | 10.6 |
| 17 | Support Structure Organization | 5 | 2 | 2 | 12 | 12 | 45 | 12 | 5 | 7 | 102 | 6.0 |
| 14 | Public & Non-Gov. Organizations | 1 | 1 |  |  | 8 | 5 | 12 | 3 | 1 | 31 | 2.2 |
| 1 | Government |  |  |  | 1 | 2 | 6 | 3 |  |  | 12 | 12.0 |
| 1 | Science Park |  |  |  |  |  |  |  |  |  | 0 | 0.0 |
|  | **Total - inlinks** | 21 | 23 | 4 | 25 | 42 | 86 | 43 | 13 | 18 | 275 |  |
|  | **Mean** | 0.9 | 0.8 | 0.4 | 8.3 | 8.4 | 5.1 | 3.1 | 13.0 | 18.0 |  |  |

## Analysis and Discussion

*York Science Park*

The infrastructure around the SP builds a network with three main brokers forming a triangle that connects together the business service and support structure organizations situated in the middle with different sub-networks (see Figure 1). Having the highest in-degree (24) and out-degree (37) the *University of York* is the most important actor and connects the central actors with spin-outs, knowledge-based companies, consulting offices, and third sector organizations. The second most important broker is the business developer *Science City York* (SCY), which acts as an intermediary between the industry and the central business services and support structure organizations. The third broker is the SP that connects the heart of the network with some firms and actors related to the development of the SPs in the UK.

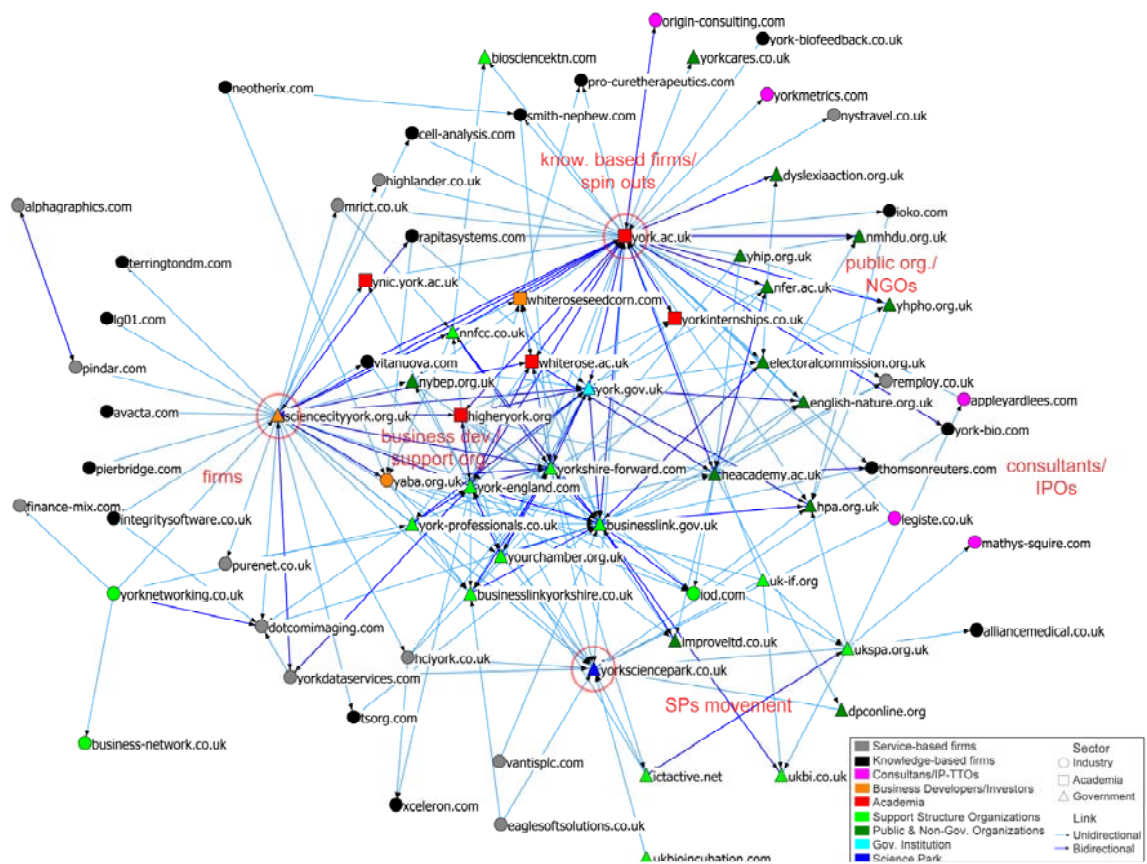

**Figure 1. Inter-linking network of York Science Park (YSP)** (A coloured version is available at: http://home.wlv.ac.uk/~in1493/issi-11/fig-1.jpg)

The analysis of the presence, or the lack of presence, of the actors listed in the framework as well as the impact or gap of their ego-networks are described below:

*University*: Being the most central actor, the University of York is directly connected to 44 (59%) of the interconnected actors and can be considered the building block of the network. The knowledge exchange with the private sector is identified through the collaboration with nine knowledge intensive companies from the Bio, Health, and IT industry and links with four University spin-offs (yorkmetrics.com; cell-analysis.com; origin-consulting.com; rapitasystems.com). The extensive interactions with the public sector are divided in two groups; one central group with nine intermediaries and support structure organizations, and another group with eleven nodes formed by independent bodies and organizations within the third sector. The enterprise initiatives of the University are also reflected through three consulting offices, two of which are spin-outs and the last being an IP office (appleyardlees.com). The presence of the University in the SP is extended through 5 other actors: the university consortium *White Rose* (whiterose.ac.uk) that promotes collaboration between the Universities of Sheffield, Leeds, and York, its seed capital investor *White Rose Technology Seedcorn Fund* (whiteroseseedcorn.com), *York Neuroimaging Centre* (ynic.york.ac.uk), *Student Internship Bureau* (yorkinterships.co.uk), and *Higher York* (higheryork.org), which is a partnership between higher education institutions to offer training and consultancy for local businesses. In addition, the central position of the three academic actors who aim to exploit the academic knowledge and technology shows how they collaborate with support structure organizations to attract public funding and create the adequate conditions for new businesses to flourish. The prominent role of the university as the main knowledge producer in this knowledge-based network confirms a two-way flow of influence between the university and the other actors, becoming the liaison between the institutional spheres and intermediaries. The SP is used as the arena to carry out outreach activities that promote an entrepreneurial involvement and identity which are critical for a sustainable economic and social development in the region (Etzkowitz, et al., 2000; Godin & Gingras, 2000).

*Research centres: York Neuroimaging Centre* (ynic.york.ac.uk) is a research facility established by the University of York to produce multidisciplinary research and serve the demands of the university, National Health Services (NHS) and the industry. However, its level of integration is very low around the SP.

*Consulting organizations:* Six actors can be identified whose main activity is consultancy but only three of them are connected to the network. The increased demand by industry and government for technology and customized solutions has led to the establishment of two spin-off companies rooted in the Department of Computer Science, namely *YorkMetrics* (yorkmetrics.com) and *Origin Consulting* (origin-consulting.com), and a third consulting organization called *Legiste* (legiste.co.uk). In addition, the links between the University departments with independent bodies and NGOs are also established by consultation and research contracts. These quasi-academic actors only link with the University, which could be due to the fact that they are likely to have their customers outside the park, and these customers would not be identified by the link analysis method used here. Another reason could be that their relationships with the customers are not strong enough to be advertised through hyperlinks – or are kept secret for commercial confidentiality reasons. On the other hand, the three consultants that do not engage in R&D activities are isolated, including electrical, franchise and educational consultancies. This shows that the University actively offers consultancy services and has formalized the commercialization of academic knowledge through spin-off companies, although the links of the consultants with potential customers are not evident.

*Technology transfer offices:* The network includes three IP & Trademark offices. *Appleyaard Lees* (appleyardlees.com), *Mathys and Squire* (mathys-squire.com), and the isolated *Murgitroyd & Company*. Although the first office works with the University, they have a peripheral role in the network without being connected with any spin-out or enterprise. The low connectivity of this group of actors might call into question the need for studying the external relationships of the actors to determine whether certain actors could be investigated with the help of hyperlinks. Nevertheless, the reason behind their low activity might be the low proportion of enterprises that apply for a patent (2.5%) or register a trademark (5.2%) in the region (BIS,

2009), and the negative effect on the licensing activities as a result of the overemphasis on the generating spinouts, regardless of their quality, driven by the availability of public funds that see the new firms as a source of employment (Lambert, 2003:58-62).

*Incubators:* Despite the lack of an incubator, the network contains many new ventures. This role is filled by the University in an active collaboration with its partners, including York SP, the business developer SCY, and the White Rose consortium. This reflects the importance of the structural collaboration between university and government to establish intermediary organizations that are capable of encouraging technology transfer and firm-formation.

*Investors: Yorkshire Association of Business Angels* (yaba.org.uk) and *White Rose Technology Seedcorn Fund* (whiteroseseedcorn.com) are responsible for injecting capital and commercial expertise into new businesses. The first brokers business angels with entrepreneurs looking for finance and mentoring. The second invests in early stage commercial opportunities based on new technology emerging from the universities of York, Leeds and Sheffield until they are ready for later stage investors. As risk capital providers, they receive funding from the RDA and academia and are closely interconnected with the University and various business supporters, while their direct links with businesses are limited to only three spin-out companies. These few ties could be the result of the often informal relationships between early-stage investors and new firms. Another important actor is the business developer *Science City York* (SCY) (sciencecityyork.org.uk), which was established by York City Council and the University of York to ensure York's economic regeneration and build a reputation as a centre of scientific and technological excellence. It works in partnership with the University and support structure organizations to stimulate the creation and growth of business through public supported mentoring, facilitating investment funds within bioscience, IT, and the creative industries, and contributing to a growing local knowledge-based industry (Lambert, 2003:72). As might be expected, SCY fosters firm-formation activities around the University and builds a bridge between companies and support structure organizations at the heart of the network. This group of actors is interconnected with key actors and shows the importance of seed-stage investments in this context as well as the difficulties to identify the ties at this stage of development. Furthermore, it illustrates the efforts of the public sector to encourage the collaboration required for an open innovation process.

*Government agencies:* There are two groups of organizations with public roots; one formed by business support organizations which occupy a central position, and another formed by third sector organizations and independent bodies which are more peripheral, on the left side of Figure 1. The few actors which are involved in the development of SPs are situated towards the bottom as well as the networking organizations. The largest group provides economic resources, specialised advice and support for the network. It is led by the support organizations *Business Link* (businesslink.gov.uk) that delivers publicly funded business support products and services designed to help new businesses, the regional development agency (RDA) *Yorkshire forward* (yorkshire-forward.com), *York England* (york-england.com) that supports the regional businesses attracting new investment for the region, and *City of York Council* (york.gov.uk). As might be expected, these public actors at the heart of the network encourage the co-operation among the nodes through partnerships and funds to key brokers, and are surrounded by the York SP, business service support organizations such as *Your Chamber*, *Higher York*, *Institute of Directors* (*iod.com*) and networking organizations ( york-professionals.co.uk; yorknetworking.co.uk; business-network.co.uk).

The second group exhibits an intensive cross-fertilization between the University and independent bodies and NGOs. This relationship is established through partnerships and research contracts which bring problems and needs into the University and at the same time allow the organizations to take advantage of the accumulated knowledge to provide efficient advice and services to policy-makers and the wider community. These actors have local offices or are based in the park and are linked to university departments that are judged as world-leading and top-ranked by the last Research Assessment Exercise (RAE, 2010), such as the Biology department, which is linked to *Natural England* (english-nature.org.uk); Health Economics and Science which is linked to *Yorkshire and Humber Public Health Observatory*, *Yorkshire & Humber Improvement Partnership*, *Health Protection Agency* and *National Mental Health Development Unit* (yhip.org.uk; yhpho.org.uk; hpa.org.uk; nmhdu.org.uk); Social Policy and Social Work, which is linked to the *National foundation for research* (nfer.ac.uk); and the department of Psychology, which is linked to the *Higher Education Academy* and *Dyslexia Action* (heacademy.ac.uk; dyslexiaacition.org.uk). Finally, there are also a few actors that work to promote the development of SPs and business incubators (ukspa.org.uk; ukbi.co.uk; uk-if.org; iactive.net). The range of outreach activities shows that the third mission of the University of York is not limited to an entrepreneurial approach, but also embraces a wide social function assisting policy-makers and the civil society. The balance between profitable and social activities in the network could be seen to give the York SP a science shop's identity (Fischer, et al., 2004). However, there are key aspects such as the secure access to public and infrastructural funding and the level of institutionalisation of the SP that distinguishes it from the traditional science shops.

*Firms:* There are 30 actors which can be identified as knowledge-based firms and 24 service-based firms in the SP. The first group exhibits a higher proportion of connectivity, having 18 (60%) firms linked to the network and the highest degree of interconnectivity with the business developer SCY and the University. On the other hand, the second group has 13 (54%) firms linked to the network. Due to the significant impact of SCY and the University among both groups of firms, 24 (77%) of the firms are linked to one or both of the actors, which are the main brokers between the public and industrial sector. Furthermore, most of the businesses are classified within the three clusters of interest (bioscience, IT, creative industries) of SCY, while nine knowledge-based companies have spun out from the university, seven from the University of York and two from Leeds. However, despite the higher level of interlinking of the knowledge-based firms, the presence of the service-based firms is also significant. This could be the result of the range of firms that are eligible for support from SCY (including bio-related, graphic and web design companies) and might be driven by the efforts of public organizations to increase the number of firms and jobs created instead of the quality of these as pointed out in the UK Life Science Start-up report (2010). In addition, the region of Yorkshire, with the highest public expenditure in the UK (Robson & Kenchatt, 2010), focuses its efforts by distributing small amounts of money to a large number of firms rather than concentrating the funding in firms with a high growth potential, hence taking a lower risk or not being able to identify promising firms. However, this financial support to a high number of service-based start-ups is unsustainable because the region is not able to attract private investments nor to form a science cluster (Nottingham BioCity, 2010).

The analysis of the network indicates that York SP provides a network characterized by the collaborative efforts between the university and local government to create conditions which allow knowledge-intensive businesses to flourish. It is interesting to observe how this is in line with the policy agenda and characteristics observed in York by the Lambert's review, which states: "The university's science park provides incubator space for new start-ups, while the council helps to link businesses with legal, financial and marketing professionals." (2003:72). As a producer of research, the university becomes the main source of firm-formation and knowledge-based services in the network and attracts both private and public actors as consumers. The functional university-government collaboration is supported by the important investments made by the RDA that fills its expected labour, as the major stakeholder in university outreach activities, especially in relation to supporting SMEs and local communities (Woollard, Zhang, & Oswald, 2007:390). Interestingly, these efforts coincide with the fact that the hands-on approach of the regional government involves the highest public investment in the UK and the Universities in the region have one of the highest budget in the UK, but at the same time the region also attracts one of the lowest levels of private investments (Lambert, 2003:65; Nottingham BioCity, 2010), as shown by the lack of involvement of the private sector in the interlinking network. This also suggests that the network might struggle to survive without public funding.

The university links the private, public and third sectors together and works in partnership with the RDA to set up important actors, namely the business developer SCY and the York SP, to obtain the external funding and services to support new businesses and thus complement its own seed capital unit. The links also suggest that the university uses the SP to capitalize its research via consultancy and research contracts, and the establishment of knowledge-based spin-outs. On the other hand, other actors such as the research centre and, the consulting and IP offices exhibit a low degree of activity within the SP. This could be the result of the inability of the hyperlinks to reflect their relationships accurately or the low engagement with the members of the network. In a previous study of an industrial SP in the region, the research centre and consultants also showed a low level of interconnectivity (Minguillo & Thelwall, submitted). Due to the low degree of connectivity of these three types of actors, it could be interesting to analyze their links outside the SP to find out if their activity is rather focused on external relationships.

Even though the injection of private investments in the infrastructure of the network is limited, the businesses take advantage of privileged links to the University and its partners as well as of a sub-network of actors dedicated to provide specialised business and financial support services. This promotes the growth of a high number of businesses in the park despite the lack of an incubator that supports this complex process, and thus highlights the key role of the hybridisation process among the actors to form a dynamic and flexible support infrastructure. This fundamental role is taken by the business developer SCY that together with two investors becomes one of the main intermediaries of public funding and services which allows spin-outs and SMEs to flourish. The impact of SCY leads to a close collaboration between the knowledge-based companies and the University, and the significant integration of service-based firms driven in part by aligning the aims of York SP and SCY to host and support companies with a wide range of profiles. Other actors taking advantage of the research and social commitment of the university are the independent bodies and third sector organizations which engage in partnerships that lead to knowledge exchange and application in the wider community, reinforcing the social function of the University of York.

**Conclusions**

In answer to the first research question, the study shows that the *SP actor framework* can be used to identify eight of the nine types of actors represented in the network. The lack of an incubator is interesting due to the fact that the incubation process seems to be essential for the formation and development of new businesses and receives considerable attention in the literature, being one of the main actors in the innovative infrastructures. Nevertheless, the dynamic infrastructure established in York SP is still able to attract and generate new businesses and university spin-offs.

In answer to the second research question, the links between the organisations coincide with the potential behaviour of only three of the actors described in the *SP actor framework*. The actors that best follow the expected patterns are the university that links all the different types of actors, the government agencies that provide the financial resources to promote the university-industry collaboration, and the investors that collaborate with the university and public business supporters to support a couple of companies with a high growth potential. On the other hand, there are actors with a low connectivity to the network such as the consulting organizations that are only connected with the university, although the consultancy services offered by the university are well represented. The knowledge-based and serviced-based firms occupy a peripheral position and are mainly connected to either the university or the business developer, respectively. Finally, the links of the research centre and the technology transfer offices do not follow the expected patterns.

However, it is important to point out that the connections should not be considered exhaustive since there might be formal and informal connections between the actors which are not represented by hyperlinks. Moreover, the strong participation of the public actors might be biased due to the need to increase government transparency and provide electronic access to government information (Jaeger & Bertot, 2010). On the other hand, the few links among the companies might be attributed to the low rate of links established by the private sector (Stuart & Thelwall, 2006; Suvinen, Konttinen, & Nieminen, 2010). Likewise, other actors such as the consulting organizations and IP offices might tend to establish few internal links, suggesting that the external environment of the SPs should also be investigated. The major limitation of the method used in this study is that it only focuses on the intra-networking dynamics of the SP while it is also important to study how they influence the region and whether they are able to extend beyond the region. Therefore, the next step will be a study of the inter-networking dynamics of the SPs, operating through internal networks among SPs and actors from different SPs in the region, as well as the external-networking dynamics, operating through external networks with the actors within the SPs.

In spite of the inherent limitations of a webometric approach, the framework makes it possible to carry out a structured analysis of the actors involved in the SPs through the identification of key actors and their expected behaviour in terms of the robustness and dynamism embedded in the SP. However, the framework still needs to be applied to other networks to determine its capacity to understand and assess these dynamic structures. Moreover, SPs are often the result of trilateral partnerships that integrate heterogeneous actors embedded in particular socio-economic conditions which influence the mission and operational procedures that make each SP a unique social environment. This makes it difficult to draw general conclusions, and more studies are needed to realize how their R&D infrastructure is configured and observe if they reflect off-line characteristics in order to determine if the interlinking networks in this context may be used as weak benchmarking indicators (Thelwall, 2004). The need for broadening the R&D indicators, the cost of data collection related to these studies (PREST/CRIC, 2006) and the broad platforms of interaction and communication in the network society (Castells, 2009) turn link analysis into an experimental indicator that can complement other proxies to assess the S&T capacity in a region through the mapping of the articulation of the knowledge infrastructure, the entrepreneurial role of the university and the degree of impact of the regional innovation strategy.